\documentstyle[aps,preprint,tighten]{revtex}

\begin{document}

\draft

\title{Casimir force on a piston}
\author{R. M. Cavalcanti\footnote{Eletronic address: {\tt rmoritz@if.ufrj.br}}}
\address{Instituto de F\'{\i}sica, Universidade Federal do Rio de Janeiro \\
Caixa Postal 68528, 21941-972 Rio de Janeiro, RJ, Brazil}
\date{January 26, 2004}

\maketitle

\begin{abstract}

We consider a massless scalar field obeying Dirichlet
boundary conditions on the walls of a two-dimensional $L\times b$
rectangular box, divided by a movable partition (piston)
into two compartments of dimensions $a\times b$ and $(L-a)\times b$. 
We compute the Casimir force on the piston in the limit $L\to\infty$.
Regardless of the value of $a/b$, the piston is attracted to
the nearest end of the box. Asymptotic expressions
for the Casimir force on the piston are derived for $a\ll b$ and 
$a\gg b$.

\end{abstract}

\pacs{PACS numbers: 03.70.+k, 11.10.-z, 11.10.Kk}


\section{Introduction}

In 1948 Casimir predicted a remarkable macroscopic 
quantum effect: two conducting and neutral parallel plates should
attract each other due to the disturbance of the
vacuum of the electromagnetic field caused by their presence \cite{Casimir1948}
(for a general review on the Casimir effect, see 
Ref.\ \cite{Bordag2001}). Inspired by that result,
a few years later Casimir suggested that the zero-point 
pressure of the electromagnetic 
field might yield the stresses postulated by Poincar\'e
in order to explain the stability of the electron \cite{Casimir1953}. 
Boyer, however, showed that the Casimir force
for a conducting spherical shell is repulsive \cite{Boyer1968},
thus invalidating Casimir's model for the electron.

Boyer's result brought attention to the fact that the 
attractive or repulsive character of the Casimir force
depends on the geometry of the configuration.
This has been investigated in detail for fields (scalar
or electromagnetic) confined in a $d$-dimensional rectangular box
\cite{Lukosz1971,Mamaev1979,Ambjorn1983,Caruso1991,Hacyan1993,Li1997,Maclay2000}.
Let us consider, for instance, 
a massless scalar field subject to Dirichlet boundary conditions at 
the walls of the two-dimensional box $0\le x\le a$, $0\le y\le b$. 
The vacuum energy is formally given by ($\hbar=c=1$)
\begin{equation}
\label{Ebare}
E_0(a,b)=\frac{1}{2}\sum_{j,k=1}^{\infty}\omega_{jk},
\qquad\omega_{jk}=\sqrt{\left(\frac{j\pi}{a}\right)^2
+\left(\frac{k\pi}{b}\right)^2}.
\end{equation}
One can perform the summation using analytic regularization (AR);
the result is (see Appendix \ref{apA})
\begin{equation}
\label{E0}
E_{0,\,{\rm AR}}(a,b)=-\frac{ab}{32\pi}\,Z_2(a,b;3)+\frac{\pi}{48}\left(
\frac{1}{a}+\frac{1}{b}\right),
\end{equation}
where $Z_2$ is an Epstein zeta function \cite{Epstein}.
An analysis of (\ref{E0}) shows that the sign of the Casimir
tension $T=-\partial{E_{0,\,{\rm AR}}}/\partial{A}$
(where $A=ab$ is the area of the box) 
depends on the ratio $b/a$: it is positive if $1\le b/a < 2.74$ and 
negative if $b/a >2.74$ \cite{Bordag2001}. 

There are,
however, at least two reasons for which one should be suspicious
of the use of Eq.\ (\ref{E0}) as the basis for such an analysis. 
First, it does not take into account the contribution to the vacuum energy 
from the region outside the box, 
which, in principle, also depends on its dimensions. (This problem
was discussed recently in \cite{Manzoni}, but the solution proposed
there is incomplete.) Second, its finiteness is an artifact of
the AR scheme: more than just regularizing integrals or sums,
it also does a certain amount of renormalization by automatically
subtracting power-law divergences (in this respect,
AR is similar to dimensional regularization \cite{DST}). 
This is precisely what happens here. If one regularizes
the sum over modes in Eq.\ (\ref{Ebare}) with a smooth cutoff
function and performs the sum using the Abel-Plana formula,
one obtains (see Appendix \ref{apB})
\begin{equation}
\label{E_AP}
E_{0,\,{\rm cutoff}}(a,b)=C_1(\Lambda)\,ab+C_2(\Lambda)\,(a+b)
+E_{0,\,{\rm AR}}(a,b),
\end{equation}
with $C_1(\Lambda)\sim\Lambda^3$ and $C_2(\Lambda)\sim\Lambda^2$ 
as $\Lambda\to\infty$. (We have discarded terms that vanish in
that limit.)

The difference between $E_{0,\,{\rm cutoff}}$ and $E_{0,\,{\rm AR}}$
would be harmless if the first two terms on the r.h.s.\ of
Eq.\ (\ref{E_AP}) could be absorbed into counterterms.
Let us forget for a moment the problem of neglecting the exterior
modes, and examine this question. The first term has the form
$\epsilon_0ab$, where $\epsilon_0$ is the energy density of
the vacuum in the absence of the box. It can be cancelled by
a ``cosmological constant'' counterterm, a constant added
to the Hamiltonian density in order to make the vacuum energy
in free space equal to zero. The problem lies in the second term:
being proportional to the perimeter of the box, it may be
interpreted as (part of) the self-energy of its walls.
Such a term {\em cannot} be eliminated by a renormalization of the
parameters of the theory \cite{Barton,Jaffe}. (This problem 
also occurs in the parallel plates configuration. In that case,
however, it can be ignored if one is interested only in the force
between the plates, for their self-energies do not depend
on the distance between them. In the present case, the dismissal of
the self-energy of the box walls could be justified if 
perimeter-preserving deformations are the only ones allowed.)

\begin{figure}[h]
\begin{picture}(500,100)(-20,0)   
\put(0,0){\line(1,0){400}}
\put(0,100){\line(1,0){400}}
\put(0,0){\line(0,1){100}}
\put(120,0){\line(0,1){100}}
\put(400,0){\line(0,1){100}}
\put(10,50){\Large{$b$}}
\put(55,10){\Large{$a$}}
\put(240,10){\Large{$L-a$}}
\put(53,48){\Large{A}}
\put(253,48){\Large{B}}
\end{picture}
\vspace{0.5cm}
\caption{Two-dimensional $L\times b$ rectangular box. A
movable partition (a piston) divides it into two compartments, A and B, 
of dimensions $a\times b$ and $(L-a)\times b$, respectively. We shall
assume that $a,b\ll L\to\infty$.}
\label{fig}
\end{figure}
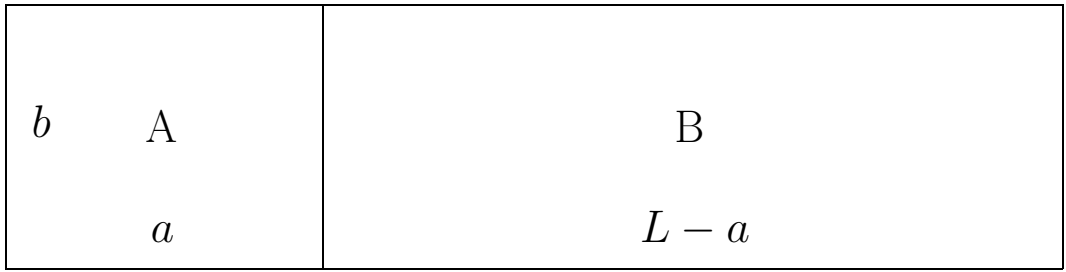

In this work we shall examine a slightly different system
in which both problems can be ignored. 
Instead of the box discussed above,
we shall consider a box of dimensions $L\times{b}$
divided by a movable partition, or piston,
into two compartments, A and B, of dimensions 
$a\times b$ and $(L-a)\times{b}$, respectively
(see Fig. \ref{fig}). 
If one is interested --- as we are --- in computing the Casimir force
{\em on the piston}, then the contribution to the vacuum energy from the region
outside the box can be ignored, as it is not affected by the position
of the piston.
In addition, as will be shown below, the divergent terms 
in the Casimir energy are naturally eliminated when one
computes the force on the piston. 
We shall compute this force in the limit $L\to\infty$ and 
show that it pulls the piston to the nearest end of the box 
regardless of the value of the ratio $a/b$. We shall also
derive asymptotic expressions for the force for 
$a\ll{b}$ and $a\gg{b}$. 


\section{Casimir force}

The total energy of the vacuum for the system described in the
previous paragraph (and depicted in
Fig.\ \ref{fig}) can be written as the sum of three terms:
\begin{equation}
\label{Etot}
E_0=E_0^{\rm A}+E_0^{\rm B}+E_0^{\rm out}.
\end{equation}
Using the cutoff regularization discussed in Appendix \ref{apB},
the first two terms are given by $E_0^{\rm A}=E_{0,\,{\rm cutoff}}(a,b)$
and $E_0^{\rm B}=E_{0,\,{\rm cutoff}}(L-a,b)$ [see Eq.\ (\ref{E_AP})],
so Eq.\ (\ref{Etot}) becomes
\begin{equation}
\label{Etot2}
E_0=E_{0,\,{\rm AR}}(a,b)+E_{0,\,{\rm AR}}(L-a,b)+C_1(\Lambda)\,Lb
+C_2(\Lambda)\,(L+2b)+E_0^{\rm out}.
\end{equation}
The Casimir force on the piston is given by $-\partial E_0/\partial a$.
Since the last three terms on the r.h.s.\ of Eq.\ (\ref{Etot2})
do not depend on the position of the piston, we obtain the following
result for the Casimir force on it:
\begin{equation}
\label{F1}
F=-\frac{\partial}{\partial a}\left[E_{0,\,{\rm AR}}(a,b)+E_{0,\,{\rm AR}}(L-a,b)\right].
\end{equation}
As anticipated, although the total vacuum energy contains divergent terms
and a term ($E_0^{\rm out}$) that one does not know how to compute,
the Casimir force on the piston is finite and can be computed exactly.

The result one obtains for the force inserting (\ref{E0}) into 
Eq.\ (\ref{F1}) is not very illuminating, so, before we actually compute $F$,
let us derive an alternative expression for $E_{0,\,{\rm AR}}(a,b)$.
In order to do that, it is convenient
to define the auxiliary function
\begin{equation}
\label{S1}
S(m,a;s):=\pi^{-s/2}\Gamma\left(\frac{s}{2}\right)
\sum_{n=-\infty}^{\infty}\left[\left(\frac{m}{\pi}\right)^2
+\left(\frac{n}{a}\right)^2\right]^{-s/2}\quad({\rm Re}(s)>1).
\end{equation}
Its analytic continuation to the complex $s$-plane
(with simple poles at $s=1,-1,-3,\ldots$)
is given by \cite{Ambjorn1983}
\begin{equation}
\label{S2}
S(m,a;s)=\frac{am^{1-s}}{\pi^{(1-s)/2}}
\left[\Gamma\left(\frac{s-1}{2}\right)+4\sum_{n=1}^{\infty}
\frac{K_{(1-s)/2}(2nma)}{(nma)^{(1-s)/2}}\right],
\end{equation}
where $K_{\nu}(z)$ is the modified Bessel function. 
Eqs.\ (\ref{S1}) and (\ref{S2}) allows us to reexpress
the Epstein zeta function that appears in Eq.\ (\ref{E0}) 
as \cite{primed_sum}
\begin{eqnarray}
Z_2(a,b;3)&=&\sum_{j,k=-\infty}^{\infty}\!\!\!\!{}^{'}\,\,
\left(j^2a^2+k^2b^2\right)^{-3/2}
\nonumber \\
&=&\sum_{j=-\infty}^{\infty}\!\!\!{}^{'}\,\,\sum_{k=-\infty}^{\infty}
\left(j^2a^2+k^2b^2\right)^{-3/2}+
\sum_{k=-\infty}^{\infty}\!\!\!{}^{'}\,\left(k^2b^2\right)^{-3/2}
\nonumber \\
&=&\frac{2\pi^{3/2}}{\Gamma(3/2)}\sum_{j=1}^{\infty}
S(\pi ja,1/b;3)+\frac{2\zeta(3)}{b^3}
\nonumber \\
&=&\frac{2\pi^2}{3a^2b}
+\frac{16\pi}{ab^2}\sum_{j,k=1}^{\infty}\frac{k}{j}\,
K_1\left(2\pi jk\,\frac{a}{b}\right)
+\frac{2\zeta(3)}{b^3}\,.
\end{eqnarray}
Inserting this result into Eq.\ (\ref{E0}) yields
\begin{equation}
\label{E1}
E_{0,\,{\rm AR}}(a,b)=\frac{\pi}{48b}-\frac{\zeta(3)a}{16\pi b^2}
-\frac{1}{2b}\sum_{j,k=1}^{\infty}\frac{k}{j}\,
K_1\left(2\pi jk\,\frac{a}{b}\right).
\end{equation}
Inserting Eq.\ (\ref{E1}) and the corresponding expression
for $E_{0,\,{\rm AR}}(L-a,b)$ into Eq.\ (\ref{F1}) and
taking the limit $L\to\infty$
we obtain the following expression for the 
Casimir force on the piston:
\begin{equation}
\label{F2}
\lim_{L\to\infty}F=\frac{\pi}{b^2}\sum_{j,k=1}^{\infty}k^2
K_1'\left(2\pi jk\,\frac{a}{b}\right),
\end{equation}
where $K_1'(x)=dK_1(x)/dx$.
Since $K_1(x)$ is a monotonic decreasing function of $x$,
it follows from Eq.\ (\ref{F2}) that
$F<0$ for all (positive) values of $a/b$;
in other words, the piston is attracted to
the nearest end of the cavity. 

It is easy to
obtain an asymptotic expression for $F$ valid for
$a\gg b$: since $K_1(x)\sim\sqrt{\pi/2x}\,\exp(-x)$
for large $x$, one may retain only the term with $j=k=1$
in  Eq.\ (\ref{F2}), thus obtaining
\begin{equation}
F\sim-\frac{\pi}{2}\,(ab^3)^{-1/2}\,\exp\left(-\frac{2\pi a}{b}\right)
\qquad(a\gg b).
\end{equation}
This result has the same form as the asymptotic expression
of the Casimir force between two plates in {\em one-dimension}
in the case of a scalar field with mass $m=\pi/b$ \cite{Ambjorn1983}.
This fact has a simple physical interpretation: when $a\gg b$ the system
becomes quasi--one-dimensional, with the field acquiring an effective mass
equal to the energy gap $\Delta=\pi/b$
due to the confinement in the transverse direction.

In order to obtain an approximation to $F$ valid for
$a\ll b$, we note that $E_{0,\,{\rm AR}}(a,b)=E_{0,\,{\rm AR}}(b,a)$,
so that we can replace Eq.\ (\ref{E1}) by
\begin{equation}
\label{E2}
E_{0,\,{\rm AR}}(a,b)=\frac{\pi}{48a}-\frac{\zeta(3)b}{16\pi a^2}
-\frac{1}{2a}\sum_{j,k=1}^{\infty}\frac{k}{j}\,
K_1\left(2\pi jk\,\frac{b}{a}\right).
\end{equation}
If, on the other hand, we still express
$E_{0,\,{\rm AR}}(L-a,b)$ in Eq.\ (\ref{F1})
according to Eq.\ (\ref{E1}), we obtain an alternative
expression for the force on the piston (in the limit $L\to\infty$):
\begin{equation}
\label{F3}
F=-\frac{\zeta(3)b}{8\pi a^3}+\frac{\pi}{48 a^2}
-\frac{\zeta(3)}{16\pi b^2}+\frac{\pi b}{a^3}
\sum_{j,k=1}^{\infty}k^2K_0\left(2\pi jk\,\frac{b}{a}\right).
\end{equation}
The last term in Eq.\ (\ref{F3}) is exponentially supressed 
when $a\ll b$, so in this case we have
\begin{equation}
\label{F4}
F\sim-\frac{\zeta(3)b}{8\pi a^3}+\frac{\pi}{48 a^2}
-\frac{\zeta(3)}{16\pi b^2}\qquad(a\ll b).
\end{equation}
If one divides both sides of Eq.\ (\ref{F4}) by $b$,
the first term on its r.h.s.\ correctly reproduces 
the Casimir tension between two infinite parallel lines a distance
$a$ apart \cite{Ambjorn1983}. The other two terms are
subdominant for $a\ll b$, and yield finite size corrections
to that result.


\section{Conclusion}

We argued in this work that the knowledge
of the vacuum energy {\em inside} a rectangular cavity is not
enough for one to calculate the Casimir force on its faces.
Two ingredients are missing in such a calculation: the
knowledge of the contribution to the vacuum energy from
the region outside the cavity,
and the proper handling of divergent terms in
the regularized expression of the vacuum energy.
We then considered a slightly different type of cavity, namely,
a rectangular box divided by a piston into two rectangular compartments. In this
case, if one is interested only in the Casimir force on the piston,
those ingredients {\em can} be neglected. In addition,
the force-on-the-piston problem has two attractive
features: (i) it is a simple generalization of the single-cavity
problem, for which results are already available in the literature \cite{Lukosz1971,Mamaev1979,Ambjorn1983,Caruso1991,Hacyan1993,Li1997,Maclay2000},
and (ii) from the experimental point of view, it is simpler
to construct a cavity with a piston than a variable-size 
rectangular cavity.
Results for the electromagnetic field in a three-dimensional
rectangular cavity with a piston will be presented elsewhere.


\acknowledgments

I would like to thank C.\ Farina for his comments on a previous
version of this paper. I also acknowledge the financial support from CNPq.


\appendix

\section{}
\label{apA}

Let us evaluate the divergent sum over modes in Eq.\ (\ref{Ebare}) 
using analytic regularization. We start with the function
\begin{equation}
\label{calE1}
{\cal E}(a,b;s):=\frac{\pi}{2}\sum_{j,k=1}^{\infty}
\left[\left(\frac{j}{a}\right)^2
+\left(\frac{k}{b}\right)^2\right]^{-s/2},
\end{equation}
which is defined for ${\rm Re}(s)>2$. As we shall see, its analytic continuation
to the complex $s$-plane is well-defined at $s=-1$, so we can
define the analytically regularized Casimir energy as 
$E_{0,\,{\rm AR}}(a,b)={\cal E}(a,b;-1)$.

In order to obtain the analytic continuation of ${\cal E}(a,b;s)$ 
it is convenient to rewrite Eq.\ (\ref{calE1}) as \cite{primed_sum}
\begin{eqnarray}
{\cal E}(a,b;s)&=&\frac{\pi}{8}\sum_{j=-\infty}^{\infty}\!\!\!{}^{'}\,\,
\sum_{k=-\infty}^{\infty}\!\!\!{}^{'}\,
\left[\left(\frac{j}{a}\right)^2
+\left(\frac{k}{b}\right)^2\right]^{-s/2} 
\nonumber \\
&=&\frac{\pi}{8}\left\{\sum_{j,k=-\infty}^{\infty}\!\!\!\!{}^{'}\,\,
\left[\left(\frac{j}{a}\right)^2
+\left(\frac{k}{b}\right)^2\right]^{-s/2}
-\sum_{j=-\infty}^{\infty}\!\!\!{}^{'}\,
\left(\frac{|j|}{a}\right)^{-s}
-\sum_{k=-\infty}^{\infty}\!\!\!{}^{'}\,
\left(\frac{|k|}{b}\right)^{-s}\right\}
\nonumber \\
&=&\frac{\pi}{8}\,Z_2\left(\frac{1}{a},\frac{1}{b};s\right)
-\frac{\pi}{4}\,\zeta(s)\left(a^s+b^s\right),
\label{calE2}
\end{eqnarray}
where $Z_p(a_1,\ldots,a_p;s)$ and $\zeta(s)$ denote the
Epstein and Riemann zeta functions, respectively.
Applying the reflection formulae \cite{Ambjorn1983}
\begin{equation}
\Gamma\left(\frac{s}{2}\right)\pi^{-s/2}\zeta(s)=
\Gamma\left(\frac{1-s}{2}\right)\pi^{(s-1)/2}\zeta(1-s),
\end{equation}
\begin{equation}
a_1\cdots a_p\,\Gamma\left(\frac{s}{2}\right)
\pi^{-s/2}Z_p(a_1,\ldots,a_p;s)
=\Gamma\left(\frac{p-s}{2}\right)
\pi^{(s-p)/2}Z_p(1/a_1,\ldots,1/a_p;p-s)
\end{equation}
to Eq.\ (\ref{calE2}) and taking $s=-1$ we obtain Eq.\ (\ref{E0}).


\section{}
\label{apB}

In this Appendix we derive Eq.\ (\ref{E_AP}) using the
Abel-Plana summation formula \cite{Whittaker},
\begin{equation}
\label{AP}
\sum_{n=0}^{\infty}F(n)=\frac{1}{2}\,F(0)+\int_0^{\infty}F(t)\,dt
+i\lim_{\varepsilon\to 0^+}\int_0^{\infty}
\frac{F(\varepsilon+it)-F(\varepsilon-it)}{e^{2\pi t}-1}\,dt.
\end{equation}
$F(z)$ is an analytic function in the right half-plane,
going to zero sufficiently fast as $|z|\to\infty$, $|\arg(z)|<\pi/2$.

In order to apply the Abel-Plana formula to the series (\ref{Ebare}),
we have to introduce a smooth cutoff function $D_{\Lambda}(z,w)$:
\begin{equation}
\label{Ereg}
E_{0,\,{\rm cutoff}}(a,b)=\frac{\pi}{2}\sum_{j,k=1}^{\infty}
\left(\frac{j^2}{a^2}+\frac{k^2}{b^2}\right)^{1/2}
D_{\Lambda}\left(\frac{j}{a},\frac{k}{b}\right)
=:\frac{\pi}{2}\sum_{j=1}^{\infty}S_j.
\end{equation}
The function $D_{\Lambda}(z,w)$ must satisfy the following conditions 
in the region ${\rm Re}(z),{\rm Re}(w)\ge 0$:
(i) it is analytic in both variables;
(ii) it is real for $z$ and $w$ real;
(iii) it vanishes sufficiently fast for $|z|,|w|\to\infty$ 
(so that the regularized series is absolutely convergent); 
(iv) it is symmetric, i.e., $D_{\Lambda}(z,w)=D_{\Lambda}(w,z)$,
and (v) $\lim_{\Lambda\to\infty}D_{\Lambda}(z,w)=1$. 
An example of such a function is given by
$D_{\Lambda}(z,w)=d_{\Lambda}(z)\,d_{\Lambda}(w)$, with
$d_{\Lambda}(z)=[1+(z+1)^2/\Lambda^2]^{-2}$. 

Applying formula (\ref{AP}) to the series $S_j$ in (\ref{Ereg}), 
we can rewrite each of them as a sum of three terms, namely,
\begin{equation}
S_j^{(1)}=-\frac{j}{2a}\,D_{\Lambda}\left(\frac{j}{a},0\right),
\end{equation}
\begin{equation}
S_j^{(2)}=\int_0^{\infty}du\,\left(\frac{j^2}{a^2}+\frac{u^2}{b^2}\right)^{1/2}
D_{\Lambda}\left(\frac{j}{a},\frac{u}{b}\right),
\end{equation}
\begin{equation}
\label{S3}
S_j^{(3)}=i\lim_{\varepsilon\to 0^+}\int_0^{\infty}\frac{du}{e^{2\pi u}-1}\left[
\left(\frac{j^2}{a^2}-\frac{u^2}{b^2}+i\varepsilon\right)^{1/2}
D_{\Lambda}\left(\frac{j}{a},\frac{\varepsilon+iu}{b}\right)-\mbox{c.c.}\right].
\end{equation}
Applying (\ref{AP}) to $\sum_j S_j^{(1)}$ yields
\begin{equation}
\sum_{j=1}^{\infty}S_j^{(1)}=\int_0^{\infty}dv\left[-\frac{v}{2a}\,
D_{\Lambda}\left(\frac{v}{a},0\right)\right]
+\frac{1}{a}\int_0^{\infty}\frac{v\,dv}{e^{2\pi v}-1}
\left[D_{\Lambda}\left(\frac{iv}{a},0\right)+\mbox{c.c.}\right].
\end{equation}
Changing the variable of integration in the first integral
to $t=v/a$ and taking the limit $\Lambda\to\infty$ in the second one we
obtain
\begin{equation}
\sum_{j=1}^{\infty}S_j^{(1)}=-\frac{a}{2}\int_0^{\infty}dt\,
t\,D_{\Lambda}(t,0)+\frac{1}{24a}\,.
\end{equation}
Similarly, application of (\ref{AP}) to $\sum_j S_j^{(2)}$ yields
\begin{equation}
\sum_{j=1}^{\infty}S_j^{(2)}=-\frac{b}{2}\int_0^{\infty}dt\,t\,
D_{\Lambda}(0,t)+ab\int_0^{\infty}du\int_0^{\infty}dv\,(u^2+v^2)^{1/2}
D_{\Lambda}(u,v)+I_{\Lambda}(a,b),
\end{equation}
where
\begin{equation}
I_{\Lambda}(a,b)=iab\lim_{\varepsilon\to 0^+}\int_0^{\infty}du\int_{0}^{\infty}
\frac{dv}{e^{2\pi av}-1}\left[(u^2-v^2+i\varepsilon)^{1/2}\,
D_{\Lambda}(\varepsilon+iv,u)-\mbox{c.c.}\right].
\end{equation}
Taking the limit $\Lambda\to\infty$ in the integral above
we obtain 
\begin{eqnarray}
\lim_{\Lambda\to\infty}I_{\Lambda}(a,b)&=&
-2ab\int_0^{\infty}du\int_u^{\infty}\frac{dv}{e^{2\pi a v}-1}\,(v^2-u^2)^{1/2}
\nonumber \\
&=&-2ab\int_0^{\infty}\frac{dv}{e^{2\pi a v}-1}\int_0^v du\,(v^2-u^2)^{1/2}
\nonumber \\
&=&-\frac{\pi ab}{2}\int_0^{\infty}\frac{v^2\,dv}{e^{2\pi a v}-1}
\nonumber \\
&=&-\frac{\zeta(3)b}{8\pi^2a^2}\ .
\end{eqnarray}
Finally, let us take the limit $\Lambda\to\infty$ in Eq.\ (\ref{S3}).
Changing the variable of integration $u$ to $t=au/jb$, we obtain
\begin{eqnarray}
\lim_{\Lambda\to\infty}S_j^{(3)}&=&
-\frac{2j^2b}{a^2}\int_1^{\infty}\frac{dt}{e^{2\pi jbt/a}-1}\,(t^2-1)^{1/2}
\nonumber \\
&=&-\frac{2j^2b}{a^2}\sum_{k=1}^{\infty}\int_1^{\infty}dt\,(t^2-1)^{1/2}\,
e^{-2\pi kjbt/a}
\nonumber \\
&=&-\frac{1}{\pi a}\sum_{k=1}^{\infty}\frac{j}{k}\,
K_1\left(2\pi kj\,\frac{b}{a}\right).
\end{eqnarray}
Collecting all pieces together we obtain Eq.\ (\ref{E_AP}),
with $E_{0,\,{\rm AR}}(a,b)$ in the form given by Eq.\ (\ref{E2}) and
\begin{equation}
C_1(\Lambda)=\frac{\pi}{2}\int_0^{\infty}du\int_0^{\infty}dv\,(u^2+v^2)^{1/2}
D_{\Lambda}(u,v),
\end{equation}
\begin{equation}
C_2(\Lambda)=-\pi\int_0^{\infty}dt\,t\,D_{\Lambda}(t,0).
\end{equation}



\begin{references}
\bibitem{Epstein} The Epstein zeta function
$Z_p(a_1,\ldots,a_p;s)$ is defined as
$$
Z_p(a_1,\ldots,a_p;s):=\sum_{n_1,\ldots,n_p=
-\infty}^{\infty}\!\!\!\!\!\!\!\!\!\!\!^{'}\,\,\,\,\,\,\,\,
\left[(n_1a_1)^2+\cdots(n_pa_p)^2\right]^{-s/2}.
$$
The meaning of the primed sum is explained in \cite{primed_sum}.
\bibitem{primed_sum} A primed
sum over the indices $j,k,\ldots$ means that the term with
$j=k=\cdots=0$ is to be omitted, i.e., 
$\sum_j^{'}a_j=\sum_j a_j(1-\delta_{j0})$,
$\sum_{j,k}^{'}a_{jk}=\sum_{j,k}a_{jk}(1-\delta_{j0}\delta_{k0})$, etc.
Note that $\sum_j^{'}\sum_k^{'}a_{jk}=\sum_{jk}a_{jk}(1-\delta_{j0})(1-\delta_{k0})
\neq\sum_{j,k}^{'}a_{jk}$.
\end{references}
\end{document}